# SEARCH FOR TWO-PARTICLE MUON DECAY TO POSITRON AND GOLDSTONE MASSLESS BOSON (FAMILON).

V.A.Andreev[1], V.S.Demidov[2], E.V.Demidova[2], V.N.Duginov[3], Yu.V.Elkin[1], V.A.Gordeev[1], K.I.Gritsai[3], S.A.Gustov[3], V.G.Ivochkin[1], E.M.Karasev[1], M.Yu.Khlopov[4], E.N.Komarov[1], S.V.Kosianenko[1], A.G.Krivshich[1], M.P.Levchenko[1], T.N.Mamedov[3], I.V.Mirokhin[3], V.G.Olshevsky[3], Yu.A.Scheglov[1], G.V.Scherbakov[1], A.Yu.Sokolov[2], Yu.P.Schkurenko[2], A.V.Stoykov[3], S.I.Vorobyev[1], A.A.Zhdanov[1], V.A.Zhukov[3] .

[1] – Russian Academy of Sciences Petersburg Nuclear Physics Institute (Gatchina)

[2] – Institute for Theoretical and Experimental Physics, (Moscow)

[3] – JINR(Dubna)

[4] – Institute of the Applied Mathematics of RAS, Center for Cosmoparticle Physics "COSMION" (Moscow)

The experimental test of possible expansion for the Higgs sector is proposed. The lepton family violation will be studied. To reach this goal we are going to carry out the search for the scalar Goldstone boson in the neutrinoless muon decay $\mu+\rightarrow e+\alpha$. The asymmetry of the muon decay near the high energy edge of Michel spectrum is to be measured. To examine previous TRIUMF data the experiment "FAMILON" is prepared at the surface muon beam of JINR (Dubna) accelerator. The setup consist of the precision magnetic spectrometer and the device for µSR – analysis.

## 1. The theoretical basing

The standard model of particle interactions is based on the principle of gauge symmetry, offering the theoretically esthetical way to introduce the intermediate bosons of fundamental particle interactions, and on the Higgs mechanism of symmetry breaking which explains the difference between the weak and electromagnetic forces. After discovery of $W^{\pm}$ and $Z^0$ bosons on LEP at CERN there were only few doubts in consistency of that part of Glashow-Weinberg-Salam theory which concerns the interaction of vector bosons with quarks and leptons. It seems that the next and probably the decisive checkout of this theory would be the detection of Higgs bosons. Really, the major property of the Glashow-Weinberg-Salam theory is renormalizability. Nobody has managed to construct a renormalized theory containing massive vector bosons without introducing Higgs fields until now. In a simple version of the theory there is only one elementary Higgs doublet. It means that after implication of Goldstone degrees of freedom to construct massive charged $W^{\pm}$ bosons and neutral $Z^0$ boson, only one observable neutral Higgs boson remains. However, there are no grounds to assume that the Higgs sector is so poor. Contrariwise, there are theoretical reasons to extend the symmetry beyond the standard model and to suppose that if scalar particles exist in general, their number can be significant.

Spontaneous breaking of global family symmetry results in the prediction of Goldstone boson(s) called familon(s). In the case of global $SU(3)_H$ family symmetrybreaking, the octet of massless familons is predicted. In the model of singlet familons (Anselm, Uraltzev, 1983), the family symmetry breaking results in the prediction of a single familon state. Familon exchange leads to effective flavour changing neutral current processes, such as $\mu^+ \to 3e$ or $K \to \mu e$. However, for small familon coupling constant $f$ the probability for such processes is of the order of $f^4$, whereas decays with free familon emission are much less suppressed, being of the order of $f^2$. Thus the familon model can find much more sensitive test in the searches for free familon emission in decays of charged leptons and quarks, such as $\mu \to e\alpha$, $s \to d\alpha$, etc. In hadronic decays the search for familons decay modes is complicated by the necessity to account for the structure of hadrons, and is restricted by the selection rules in the hadron transitions. So, the $K \to \pi\alpha$ decay, on which severe experimental constraints exist, can not take place for pseudoscalar $\alpha$, and in the search for this decay the existence of familon with pseudoscalar coupling to fermions is elusive. Familon decays of charged leptons take place both for scalar and pseudoscalar couplings and are much easier interpreted in the terms of bare familon interaction. In the models with the broken local gauge family symmetry the gauge bosons appear, mediating flavour changing neutral current (FCNC) interactions. To escape the severe experimental restrictions on such interactions these models assume very large scale $F$, of family symmetry breaking ($F \gg 10^2$ GeV) and correspondingly large mass of FCNC intermediate bosons. It leads to the probability of FCNC decays for fermions with the mass $m_f$ to be of the order of $\sim m^5/F^4$, strongly suppressed relative to the probability of the ordinary weak decays, being of order of $\sim m^3/\Lambda^4$, where $\Lambda \sim 10^2$ GeV is the scale of electroweak symmetry breaking. It turns out that, though the family symmetry is local, the realistic choice of the gauge symmetry breaking turns to provide the existence of additional broken global symmetry, what leads to the existence of familon with the coupling $f \sim m_f/F$ in these models. So in the models of local family symmetry familon is also predicted and free familon emission modes with the probability $\sim m_f^3/F^2$ can serve as much more sensitive tool to probe the physics of lepton family violation, than the FCNC processes.

In the familon model, the massive neutrino can decay due to neutrino coupling with familons $\alpha$ to the channel $\nu_h \to \nu_L \alpha$, where $\nu_{h(L)}$ denotes the heavier (lighter) massive neutrino, respectively. The existence of such channel with the probability $m_\nu^3/F^2$, the scale $F$ being the scale of the family symmetry breaking, plays important role in the dark matter cosmology. Primordial neutrinos with the mass of some tens eV are popular dark matter candidates. Familon decays can lead to neutrino lifetime less than the age of the Universe, giving rise to scenarios of structure formation by unstable massive neutrinos. The neutrino lifetime is constrained in these scenarios: from below – by the condition of structure formation and from above – by the condition of slowing down of evolution of the structure after formation. The former condition provides the formation of the structure before neutrino decay. The latter condition is needed to remove the most part of dark matter from the structure so that the structure evolves much slower and survives to the present time. These conditions fix the scale $F$ of the family symmetry breaking and adjust the predictions for familon decays of charged leptons. In 1985 Anselm, Uraltsev and Khlopov [1] have estimated the scale of family symmetry breaking from the unstable neutrino models of structure formation and predicted the rate of $\mu \to e\alpha$ decay close to the level of sensitivity, reached in TRIUMF. To make more exact the experimental situation in this field like as the *experimentum crucis* for dark matter cosmological scenarios and familon models, underlying them. This aspect of searches for rare decays was strongly evolved in the successive development of cosmoparticle physics, studying fundamental relationship between cosmology and particle physics.

Cosmoparticle physics suggests new links in the relationship between cosmology, astrophysics and experimental physics, and, in particular, new motivations for nuclear and particle physics experiments [2]. In particular, it embeds observational astronomy and experimental particle physics into joint crossdisciplinary studies, putting into definite correspondence observational programs and specific nuclear and particle physical experimental research. In addition to widely known direct experimental searches for cosmic fluxes of dark matter particles (underground WIMP searches) new nontrivial links between cosmological problems and physical experiments appear in the context of cosmoparticle physics. Having some definite physical model as the basis for inflationary cosmology with baryosynthesis and dark matter, one may fix the parameters of the considered model from the condition of its consistency with cosmological and astrophysical data, thus making definite predictions for new particle physical processes. Experimental search for such processes in the context of cosmoparticle physics plays the role of experimental test for considered dark matter cosmology. Self consistent description of formation of the observed structure in big bang Universe, accounting for the observed isotropy of microwave background radiation appeals to the existence of nonbaryonic dark matter. The physical nature of dark matter particles is usually found in the so-called "hidden" sector of particle theory (axions, massive neutrino, SUSY particles – neutrallinos etc). Both theoretical arguments and observational data seem to favor mixed scenarios of the large-scale structure formation, which somehow combines properties of hot, cold and unstable dark matter. However, in general, dark matter candidates follow from different physical motivation. The parameters of dark matter are free and in the most cases inaccessible for direct experimental determination. That is why, the particle models are of special interest, predicting different dark matter candidates from the same theoretical grounds and relating their parameters to such phenomena of particle physics, which are accessible to experimental search. One of the interesting examples of such scheme is given by the theory of broken family symmetry [2,3]. This theory, recently evolved into the form of model of horizontal unification, offers the possibility to use the set of predictions for family violation decays as experimental probe for physical cosmology, elaborated in its framework.

The theory of broken family symmetry [2,3] is based on the standard model of electroweak interactions and QCD, supplemented by spontaneously broken chiral local $SU(3)_H$ symmetry for quark-lepton families. The set of horizontal Higgs fields, $\chi^v_{\alpha\beta}$ (n=0,1,2), is necessary to break $SU(3)_H$, where $\alpha, \beta=1,2,3$ – $SU(3)_H$-indices. Though there is the only electroweak Higgs doublet present, the theory provides natural inclusion of Peccei-Quinn symmetry $U(1)_H \equiv U(1)_{PQ}$, being associated with heavy Higgs fields and it gives natural solution for QCD CP-violation problem. The masses of quarks and leptons in this model are induced by their "see-saw" mixing with hypothetical heavy fermions, having direct Yukawa coupling with horizontal Higgs. The mass hierarchy between families appears to be inverted with respect to the hierarchy of symmetry breaking, which breaks according to following scheme:

$$SU(3)_H \otimes U(1)_H \xrightarrow{\varsigma^0} SU(2)_H \otimes U(1)_H \xrightarrow{\varsigma^1} U(1)_H \xrightarrow{\varsigma^2} I .$$

The global $U(1)_H$ symmetry breaking results the existence of axion like Goldstone boson - which has both diagonal and nondiagonal couplings with quarks and leptons

$$L_{\alpha\beta} = \alpha g_{\alpha\beta} \overline{f_\alpha} (\sin\vartheta_{\alpha\beta} + i\cos\vartheta_{\alpha\beta\gamma_5}) \overline{f_\beta} + h.c.$$

where $f_1, f_2, f_3 = u, c, t; d, s, b; e, \mu, \tau$; $g_{\alpha\beta}, \vartheta_{\alpha\beta}$ –the parameters of the model.
Additionally, Goldstone boson $\alpha$ is simultaneously invisible axion, familon and majoron, named archion [4]. The couplings of the $SU(3)_H$ gauge bosons due to the chiral assignment of fermions have the form:

$$L_{int} = \frac{1}{2} g_H H^A_\mu \left\{ \overline{f_L} \gamma^\mu \lambda_A f_L - \overline{f_R} \gamma^\mu \lambda_A^T f_R \right\} ,$$

where the $\lambda_A$ are the standard Gell-Mann matrices and $g_H$ is a gauge constant.

The mass spectrum of the bosons strongly depends on Higgs sector of the model [5]. Alongside with prediction, that archion shares properties of invisible axion, the considered model predicts the existence of neutrino mass for various generations with following hierarchy:

$$m_{\nu_e} : m_{\nu_\mu} : m_{\nu_\tau} \cong m_e : m_\mu : m_\tau .$$

Absolute values of neutrino masses are defined by main parameter of TBFS, by namely the scale of horizontal symmetry breaking $V_H$. This value also defines the hierarchy of the lifetime of neutrino relative to decays: $\tau \to \nu_\mu \alpha$, $\mu \to \nu_e \alpha$, in case if there are $\zeta$ sextets. Thus, relative contribution of massive neutrino and archions into cosmological density, and its time dependence, is defined by the main parameter $V_H$. The value of this parameter actually specifies the concrete model of dark matter. The approach of cosmoparticle physics, providing the formulation of system of nontrivial cosmological, physical and astrophysical restrictions, leads to the following allowed ranges of scales:

$V_H \equiv V_6 \approx 10^6$ GeV    and    $V_H \equiv 3 \cdot 10^9$ GeV $\div$ $1.5 \cdot 10^{10}$ GeV.

TBFS in variants $V_6$ and $V_{10}$ was called in [4] the model of horizontal unification (MHU). Each of the two ranges $V_H$, in the MHU, corresponds to specific dark matter model [6]. Moreover, the account for the possibility to incorporate in MHU physical mechanism of inflation and baryosynthesis leads to physical selfconsistent cosmological scenario based on MHU. The experimental test of the MHU in its low energy case $V_6$ illustrates the main features of experimental physical cosmology. Having fixed the parameters of dark matter model by its cosmological relevance, its accelerator test is possible in searches for particle physics effects, corresponding to these parameters. The Lagrangian (1.3.2) induce the following decays of charged leptons: $\mu \to e\alpha$, $\tau \to \mu\alpha$, $\tau \to e\alpha$. The predicted branching ratios for these decays are as follow [5, 7]:

$$B_\tau(\tau \to \mu\alpha) = 3.5 \cdot 10^{-3} D_{\tau\mu}^2 \Delta^2$$

$$B_\tau(\tau \to e\alpha) = 1.7 \cdot 10^{-5} D_{\tau e}^2 \Delta^2$$

$$B_\tau(\mu \to e\alpha) = 1.0 \cdot 10^{-6} D_{\mu e}^2 \Delta^2 ,$$

where $D$ are the combination of angle parameters of leptons mass matrix and $\Delta = 10^6$ GeV/$V_H$. In the considered model masses and lifetimes of dark matter particles are defined by the scale $V_H$, and by the angular parameters of matrixes and $P$. The parameters $y$ and $p$, in their turn, are connected with the angular parameters of the mass matrix of leptons. So the experimental search for rare decays make it possible to define in principle all the characteristics of dark matter particles in HDS, and to test the respective cosmological scenario, based on MHU.

## 2. Experimental method

The differential rate of $\mu \to e\alpha$ -decay is given by the expression [8]:

$$d\Gamma(\mu \to e\alpha) = \Gamma_0(\mu \to e\alpha)\left[1 - \vec{P}_\mu \vec{P}_e + 2(\vec{P}_e \vec{n})(\vec{P}_\mu \vec{n})\right]\frac{\sin\vartheta d\vartheta}{4}$$

where $\Gamma_0(\mu \to e\alpha)$ - is the full width of $\mu \to e\alpha$ - decay of an unpolarized muon, $\vec{P}_\mu, \vec{P}_e$ – polarizations $\mu$ and e, $\vec{n}$ – momentum direction of the positron.

The relative probability of the decay μ→eα to usual decay $\mu \to e\nu\bar{\nu}$ is determined by the ratio:

$$R_\alpha = \frac{\int_0^\pi d\Gamma(\mu \to e\alpha)}{\int_0^1 \int_0^\pi d\Gamma(\mu \to e\nu\bar{\nu})} = \frac{\Gamma_0(\mu \to e\alpha)}{\Gamma_0(\mu \to e\nu\bar{\nu})} .$$

The relative contribution of the decay μ→eα to usual decay in the narrow interval Δ on the spectrum edge:

$$B_\alpha = \frac{\int_0^\pi d\Gamma(\mu \to e\alpha)}{\int_{1-\Delta}^1 \int_0^\pi d\Gamma(\mu \to e\nu\bar{\nu})} .$$

Apparently, the events of μ→eα decay produce a narrow peak on the edge of the decay positron energy spectrum of usual $\mu \to e\nu\bar{\nu}$ decay, and it may be observed.

The first experimental evaluation of the magnitude $R_\alpha = \Gamma(\mu \to e\alpha)/\Gamma(\mu \to e\nu\bar{\nu})$ was obtained by Jodidio et al. [9] in the experiment at TRIUMF:

$$R_\alpha < 2.6 \cdot 10^{-6} .$$

However, the direct observation of the peak from the decay μ→e⁺α is impossible. But it is possible to study the decay $\mu^+ \to e^+\alpha$, where the absolute measurements are substituted by relative.

Actually, let us consider decays $\mu^+ \to e^+\alpha$ and $\mu^+ \to e^+\nu\bar{\nu}$ from the point of view of the positron angular distribution relative to the muon spin direction. It is clear that in case of the decay with emission of familon we have isotropic distribution of decay positrons, while the standard decay mode has strong asymmetry of positrons relative to the spin direction of muon, which is due to parity violation. Taking into account the finite capture angle and the detection energy range Δ, we have the following numbers of high energy positrons emitted along ($N^+$) and opposite ($N^-$) to the muon spin direction on the muon decay $\mu \to e\nu\bar{\nu}$:

$$N^\pm = \int_{1-\Delta}^1 \int_0^1 \Gamma_0(\mu \to e\nu\bar{\nu})[(3-2\varepsilon) \mp (1-2\varepsilon)P_\mu \cos\vartheta]\varepsilon^2 d\varepsilon \sin\vartheta d\vartheta .$$

Hence the asymmetry factor of high energy decay positrons $\mu \to e\nu\bar{\nu}$ will be:

$$C' = \frac{N^+ - N^-}{N^+ + N^-} = \frac{P_\mu}{2}(1-2\Delta)(1+\cos\vartheta) .$$

For the decay μ→eα we have due to the same reasons

$$N^+ = N^- = \frac{1}{2}\Gamma_0(\mu \to e\alpha)(1-\cos\vartheta) ,$$

and due to such process, the observed asymmetry factor of the μ→e-decay is

$$C'' = C' \frac{1}{1+R_\alpha/(2\Delta)} ,$$

where $C'$ and $R_\alpha$ were defined above. One can see that the ratio $C''/C'$ is independent of θ. This fact is important for accumulation of experimental statistic, because one can use wide-aperture detectors.

### 3. The experiment "FAMILON" at JINR

The experiment conception was elaborated in PNPI RAS [10]. The research of two-particle muon decay into positron and massless particle (familon) is reduced to the analysis of the positron spectrum in its high-energy region for the main decay mode μ → eνν. The direct observation of the monochromatic line of the rare decay μ → eα above the background decay μ → eνν not only requires a magnetic spectrometer with high energy resolution, but it is connected with absolute measurement difficulties (background events, scattering and others) and so is extremely problematic. Many difficulties may de avoided if anyone investigates decays of polarized muons and analyses energy- and angle distributions for positrons. Fig.1 presents the positron angle distribution for the angle between the positron momentum and the muon spin for the main decay and the decay under study. For the first case the distribution is isotropic, in the second case the clearly expressed asymmetry connected with parity violation exists for positron momentum with respect to muon spin. Therefore the background, i.e. the main decay mode, is much less for positrons going out in the direction opposite the muon spin compared with any other direction.

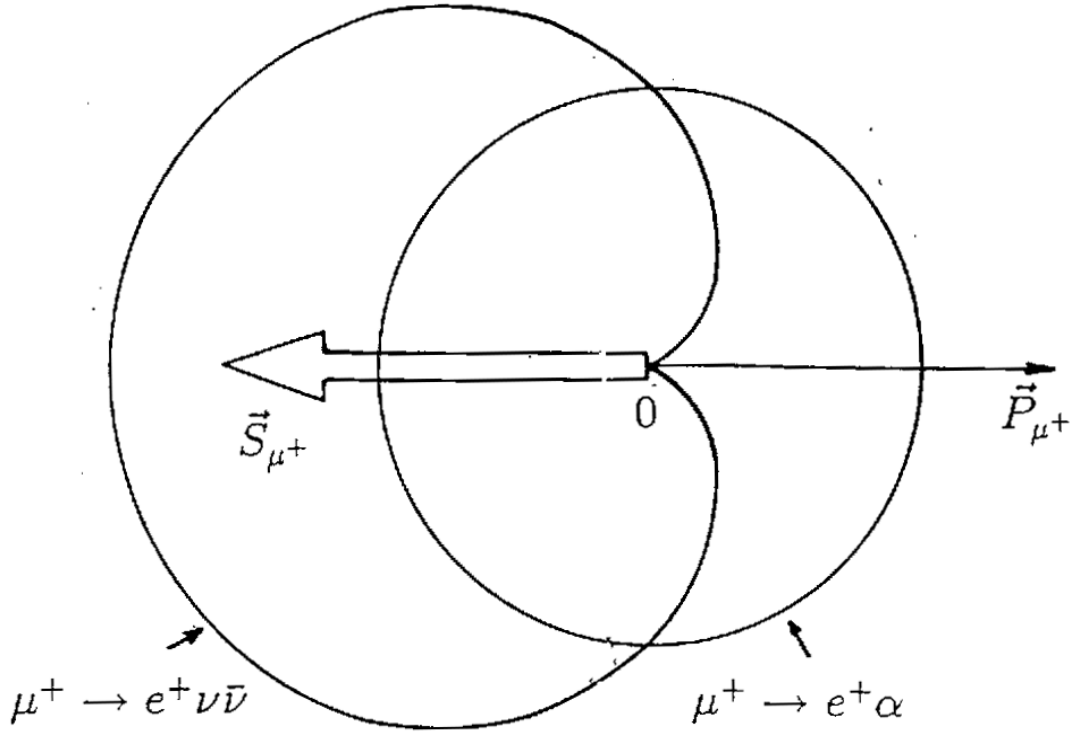

Fig.1. The angle distribution of the positrons relative to the muon spin in the decays
μ→eνν and μ→eα.

The idea to analyze at the same time the energy- and angle distributions in the experiment is realized in the following way. The positron momentum is measured by the precision magnetic spectrometer with the energy resolution of $10^{-3}$; the standard μSR- technique [11] is applied for the analysis of μ-spin –positron-momentum correlation. Spin precession spectra for polarized muons being stopped in the target with high density of the conductive electrons (to prevent depo-larization due to muonium production) are analysed as the positron energy function. If the decay $\mu \to e\alpha$ exists then the asymmetry coefficient sharply decreases near the high edge of the Mishel-spectrum. It must be noticed that systematic uncertainties are reduced to minimum in such experiment because μSR-spectrums are registered simultaneously. The layout of the experiment is presented in fig. 2. The main elements of the setup "Familon" are described in [10].

The setup is placed at the "surface" muon beam with the kinetic energy of 4 MeV produced in the decays of $\pi^+$- mesons stopped in the surface layer of the meson-generating target. In the top part of fig.2 there is the system of magnetic elements for μSR-analysis of the decay $\mu \to e\alpha$. The system has two Gelmgolts-ring couples to create cross magnetic fields: $B_\perp$ = 0-500 Gs, $B_{||}$ = 0-1000 Gs, – and 3 pairs of squared coils to compensate dissipate magnetic fields in the target region. The poles of the spectrometer magnet have the diameter of 80 cm and split of 24 cm. Magnet field in the center of the magnet is 3 kGs.

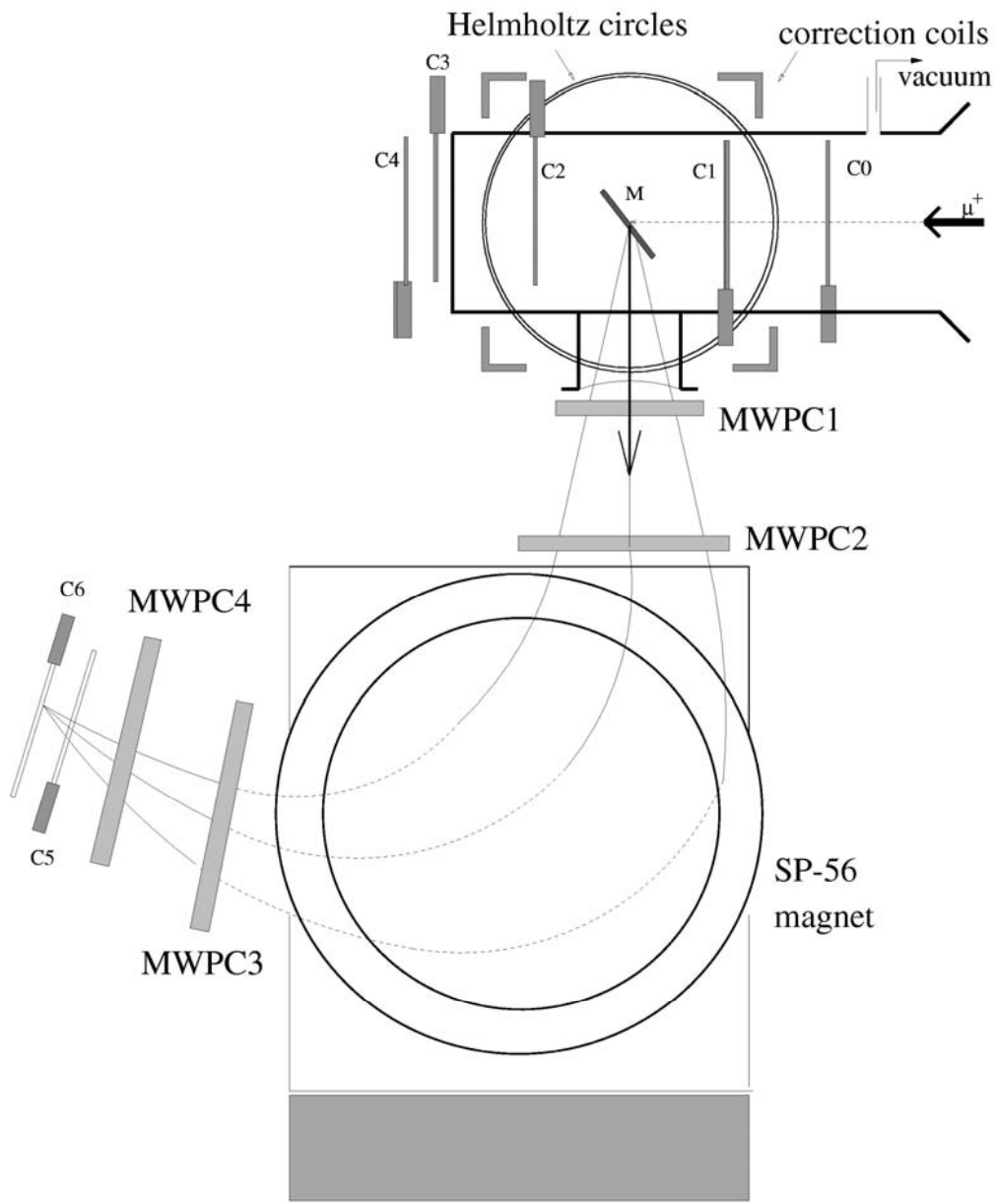

Fig.2. The layout of FAMILON set-up.

Coordinate-detecting part of the magnetic spectrometer consists of scintillation counters C1 – C6 and proportional chambers PC1 – PC4. Counters are included into the trigger system for the quick (nanosecond) selection of the events with needed configuration. The coordinate system is right, axis X is directed along the average positron momentum, axis Z is vertical. Y-, Z- planes of the proportional chambers have the step of 2 mm, split 4 mm and various channel numbers according to the chamber size. Each chamber PC1, PC2 has the size of 200×200 mm and 192 registration channels. Each chamber PC3, PC4 has the size of 400×400 mm and 384 registration channels. For chamber operation the magic mixture Ar(75%) + $C_4H_{10}$(24.7%) + freon (0.3%) was used, being prepared in the special apparatus providing the quantity of up to 200 $cm^3$/min.

As the above-mentioned results of the experiment simulation show, the existing for today equipment of the "Familon" set-up allows in principle to determine the positron energy in μ-decays with the relative precision of $2.5×10^{-3}$. But the precision of measurement of positron energy is determined both by the energy precision of the magnet spectrometer and the uncertainty of the positron energy loss in the working target, where muons stop. It is proposed in the experiment to use Al with thickness from 150 mg/$cm^2$ up to 250 mg/$cm^2$ (the average thickness is 200 mg/$cm^2$) as the working target. If the target is a single Al plate, then we cannot define the stopping point of muon in the target and therefore cannot define the energy loss for positron from muon decay. The average energy loss in the matter for relativistic positron is 2 MeV per 1g/$cm^2$. Then the uncertainty in the positron energy is 0.4 MeV and the relative precision in the positron energy at the high-energy region of the Mishel-spectrum is not better than $8×10^{-3}$. The obvious way to remove the above-mentioned uncertainty is to use the active target, which consist of some thin metal foils working as plane-parallel avalanche counter.

The prototype of the active target for the Familon experiment has been worked out and produced in PNPI RAS [12]. Investigations at the surface muon beam of the LNP JINR phasotron show high registration efficiency for the muon passage through the gas gap of the avalanche detector (99% for 1 mm $CO_2$ gap at the atmosphere pressure). Also they show the enough high ability for the selection of muons and electrons by means of the detector signal amplitude.

The electronics of the setup "Familon" is carried out in CAMAC standard and has program manipulation through the personal computer of Pentium type. The using interface means – controllers of A2-type, branch drivers and the direct connection of the computer interface with K16 controller – permit to work with up with crates disant up to 100 m. At the same time the single operation takes 30 mks, what in our experiment does not limit the rate of the statistic accumulation, because the maximum intensity of useful events is 100 $s^{-1}$.

The electronics of the proportional chambers is worked out as modules with 32 channels in each to read the chamber state. The singularity of the system for the receiving and keeping the information is the uninterrupted work after any registration circle up to the next trigger signal. Signals from the each wire go to the 16-channels RAM . The address of the channel in RAM is changed with frequency of 25 MHz. Thus in each moment the system contains the information about 16 states for each wire and these states are separated by 40 ns time interval from each other, i.e. the RAM memory time is 640 ns. The trigger signal interrupts the circle record and the information from the wire proportional chambers is read out and passed to the computer. Then the circle record to RAM cells renews up to the next trigger signal.

At present the new system (CROS3) is being elaborated in PNPI RAS for the readout of the information from coordinate detectors in the experiment "Familon". CROS3 is the third generation of the data readout system for the coordinate detectors. System CROS3 takes into account all particularities of the previous systems and develops them on the base of new technologies. The using of special micro-schemes (ASIC) and large micro-schemes with programmed logical connections (FPGA) lets to essentially decrease the system sizes and the consumed power and to rise the rate of the information processing. New standards (LVDS, PCI, Optical Link) let to reach high integration not only for the reading system but for the whole experiment electronics. The main characteristics of the system CROS3 are the following:
- program regulation of the threshold, of the delay time and of the "window"-time, where signals are registered;
- small system sizes and low dissipated power;
- interfaces in standards LVDS, Optical Link, PCI, Ethernet;
- data readout rate up to 160 Mbite/s;
- reduction of the number of cables;
- small dead time;
- the measuring of the hit channel time distribution inside the registration "window".

Software ON-LINE is realized in the system WINDOWS, and all technologies of the object-oriental programming are maintained. Software ON-LINE functions are the following [13]:
- the test of the subsystems of the setup and of the whole system;
- the operation of information accumulation in experiments;
- the management of the on-line processing in experiments;
- the parameter selection and the display of the experimental information and on-line results;
- the selection on basis of parameters and viewing of information in off-line regime;
- the viewing of tracks and of the processing results in on-line and off-line regimes.

The resultant rate of the apparatus and software is 300-1000 event/s and depends on the number of hit wires.

The main purpose of the off-line programs is to restore the positron energy using the data from proportional chambers and to create the resulting data array in the format NTUPLE for the statistical and physics processing [14].

The computer support of the experiment "FAMILON" represents the combination of the INTERNET connected computers of the institutes – members of the experiment, while the information is concentrated in PNPI. Each individual computer must have the same set of specialize programs for processing of the experimental information. The "FAMILON" set-up is mounted at the surface muon beam of the phasotron in the LNP JINR [15].

The beam particles slightly slowed down were separated from the positron admixture by the momentum in the homogeneous magnetic field. The ratio of the number of muons and positrons in the region of the maximum muon exit is 1:3. At the beam momentum of 21 MeV/c, corresponding to the maximum intensity, the momentum uncertainty is 9.5%, muon intensity is $4\times10^3$ $m^{-2}s^{-1}mkA^{-1}$, the beam sizes (half-width at half maximum) are 70 mm in horizontal and 80 mm in vertical, the positron admixture is near 200%.

The test of "FAMILON" set-up has been carried out at the accelerator test run. Using the results of the run [16] and the computer simulation by means of GEANT 3.21 the technical assignment was formulated for the completing of the setup with the additional apparatus to provide the higher accuracy in the positron energy measuring.

## 4. The simulation of the setup "FAMILON"

As it was mentioned, the method for the search of the neutrinoless muon decays is reduced to the energy dependence measuring of the asymmetry of the positron angle distribution in the muon decays near the upper edge of Mishel-spectrum ($E_m$ = 52.83 MeV). The specificity of the experiment is to measure the momentum of soft positrons in the magnetic spectrometer. It puts on some restrictions on the setup configuration.

The aim of the simulation [17, 18] is to evaluate the influence of the matter (the air, wires, scintillators and so on) on the precision of the measured positron momentum and to optimize the position of detectors in the setup in order to reach maximum energy accuracy and positron registration efficiency. Magnet sizes, the map of the magnetic field, the construction of the proportional chambers blocks, relative position of the magnet and the target – those are the base initial data having been introduced to the simulating program.

The uncertainty of the positron momentum $p$ is mainly determined by the uncertainty in reconstruction of it's direction according to hit wires in the proportional chambers in front of the magnet and back of it. It depends on the wire step size $h$ in the coordinate planes and on the range $X_n$ between chamber planes. The value $\sigma_p/p$ is used as the relative uncertainty, where $\sigma_p$ is r.m.s.- deviation. The $X_n$ – range dependence of the direction uncertainty for 52.83 MeV positron in vacuum and for the step h = 2 mm is shown in fig. 3 by falling dashed line. In the real experiment the positron multiple scattering in the matter of the setup is the main factor, which increases the errors. The mean multiple scattering angle as a function of the path length for positrons in the air is shown with asterisks connected by growing dashed-dotted line (fig.3). The summary error as mean geometrical value of these two components is shown by solid line with wide maximum at 10-18 cm. The distance between proportional chamber blocks in the air in front of the magnet and back of it must obviously be ~20 cm to get the maximal momentum measurement precision.

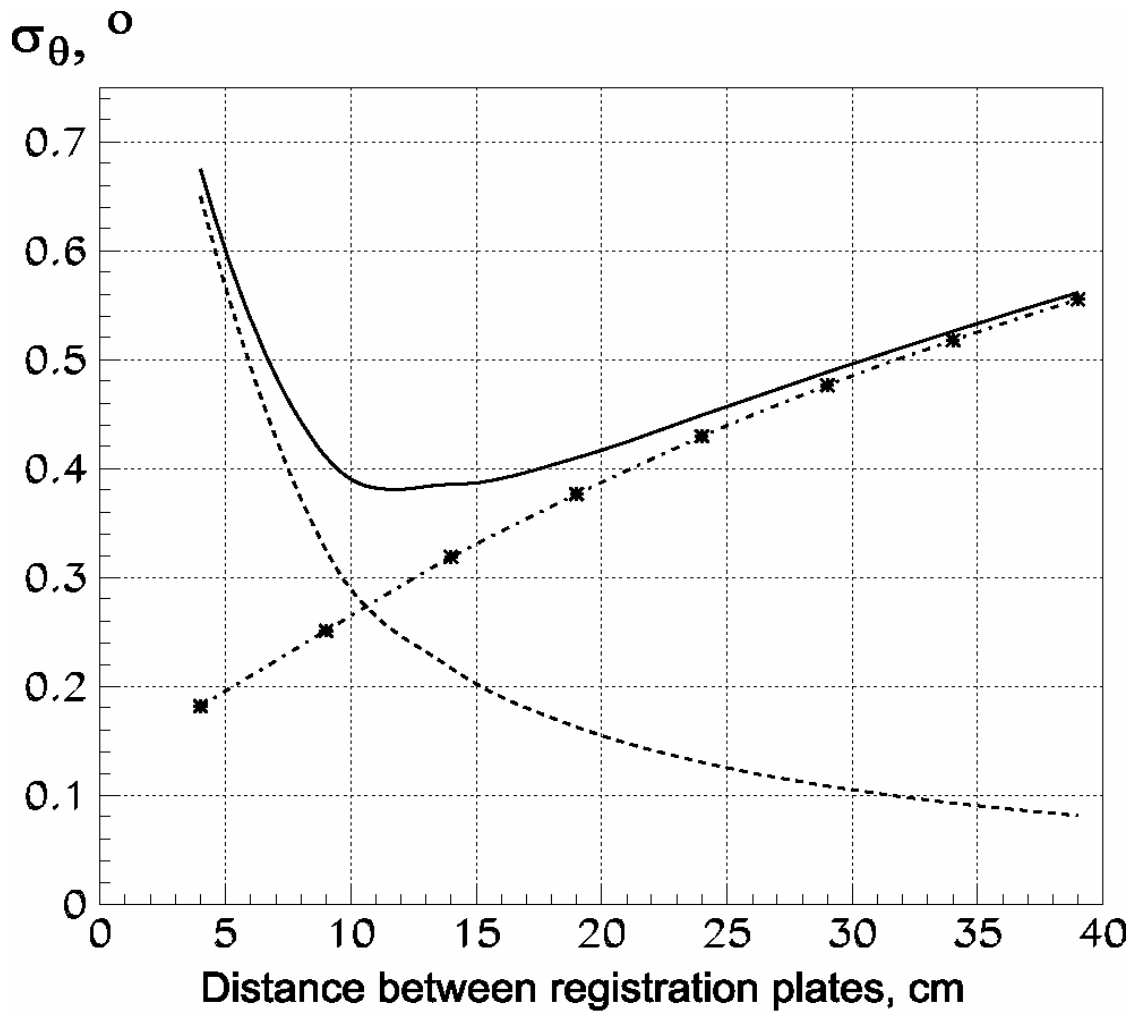

Fig.3. The dependence of the uncertainty in positron direction measurement on the distance $X$ between registration planes for positron energy $E_m$ = 52.8 MeV. Dashed line – measurement precision in vacuum; dashed-dotted line – the uncertainty because of the multiple scattering; solid line – the error in the real experiment

To evaluate the top precision of the positron momentum measurement the simulation of the monochromatic positron registration in chambers was carried out without any substance on the positron path: for 52.8 MeV (maximum in the Mishel-spectrum) and 51.8 MeV (1 MeV less). The difference in Y-coordinates on the PC4-plane for the above two energies is 37 mm. From here the transition coefficient from the coordinate error to the energy error follows, it is 0.027 MeV/mm.

To evaluate the matter influence on the momentum uncertainty the registration models were simulated for monochromatic positrons with maximal energy in the Mishel-spectrum and various setup configurations. Fig.4 shows the Y-coordinate distributions at PC4-plane for 3 cases: a) only the construction materials of the proportional chambers are taken into account and the setup is in vacuum (solid line); b) vacuum is only inside the magnet (dots); c) the whole setup is in the air (dashed line). One can see that the air puts the main contribution to the coordinate dispersion.

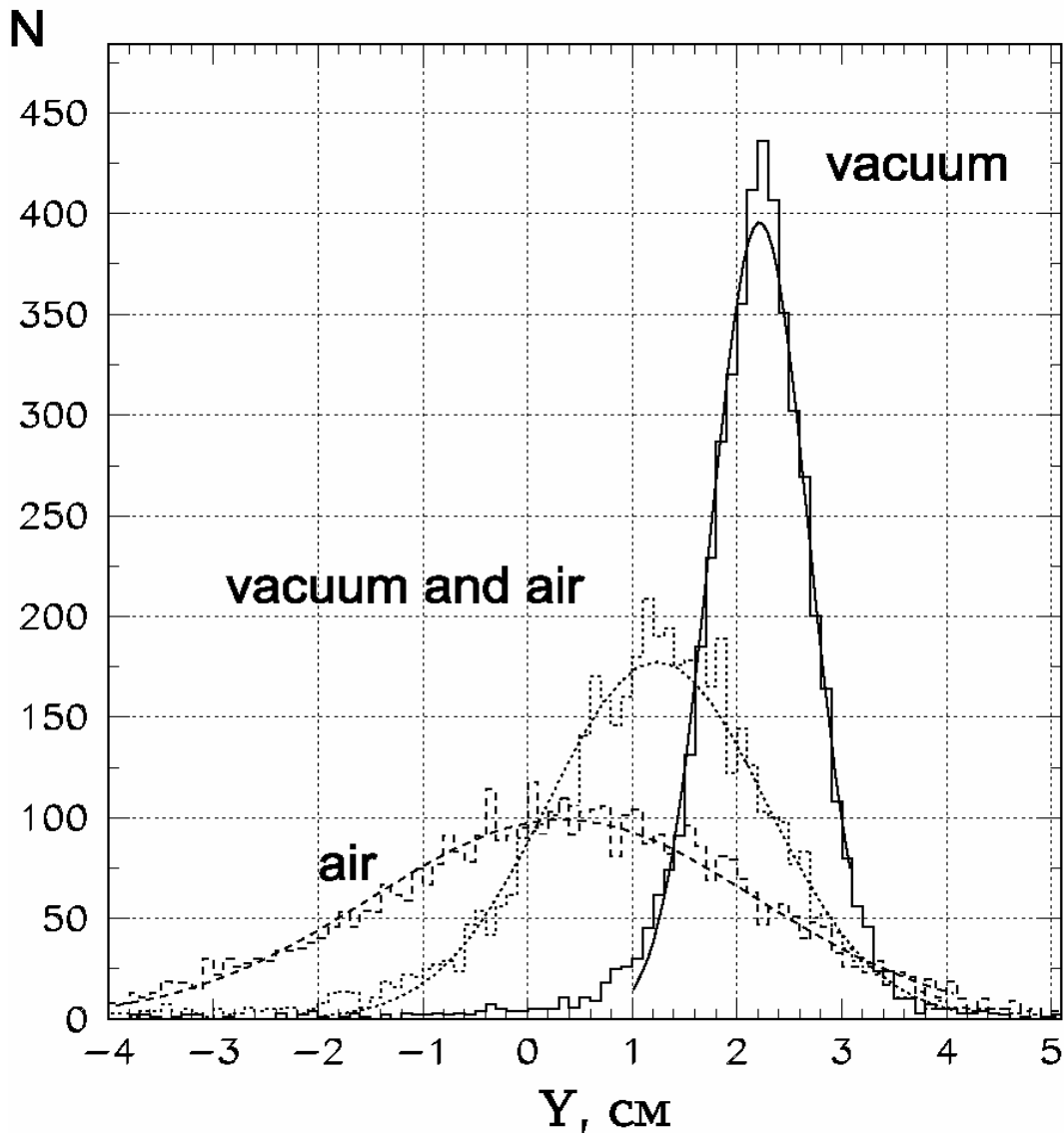

Fig.4. Y-coordinate distributions of positrons in the chamber PC4 for 3 cases:
- 3 chambers (PC2, PC3, PC4) and magnet are in vacuum (solid line);
- vacuum is only inside the magnet (dotted line);
- the whole setup is in the air (dashed line).

Analogous calculations were carried out for another 2 cases: all space between chambers is filled by helium; vacuum is inside the magnet while the rest space of the setup is in helium.

Analyzing the simulation results one can get the evaluations of the measuring precision for various configurations of the set-up. At the same time the etalon accuracy in the momentum measurements is the r.m.s.-deviation in the Y-coordinate distribution in PC4 plane (horizontal coordinate where positron hits the chamber PC4). The most important results are the following.

The top precision of coordinate measurements with step $h = 2$ mm in the ideal case (positron in vacuum) is $\sigma_p = 0.025$ MeV or $\sigma_p/p \approx 5 \times 10^{-4}$ for the maximum of Mishel - spectrum.

The maximal attainable precision for the proportional chambers of given construction (a quantity of substance along the positron path) is $\sigma_p/p \approx 2.5 \times 10^{-3}$ if the whole setup is placed in vacuum.

If vacuum is only inside the magnet then the momentum uncertainty rises up to $\sigma_p/p \approx 5 \times 10^{-3}$. If the whole registering apparatus of the setup is placed in helium the precision improves up to $\sigma_p/p \approx 3 \times 10^{-3}$.

In the case when all apparatus is in the air $\sigma_p/p \approx 9 \times 10^{-3}$.

The configuration with vacuum inside the magnet while the rest space is filled in the helium seems to be optimal. In this case $\sigma_p/p \approx 2.6 \times 10^{-3}$. Fig.5 demonstrates the trajectories of 50 positrons with energy of 52.8 MeV for this configuration.

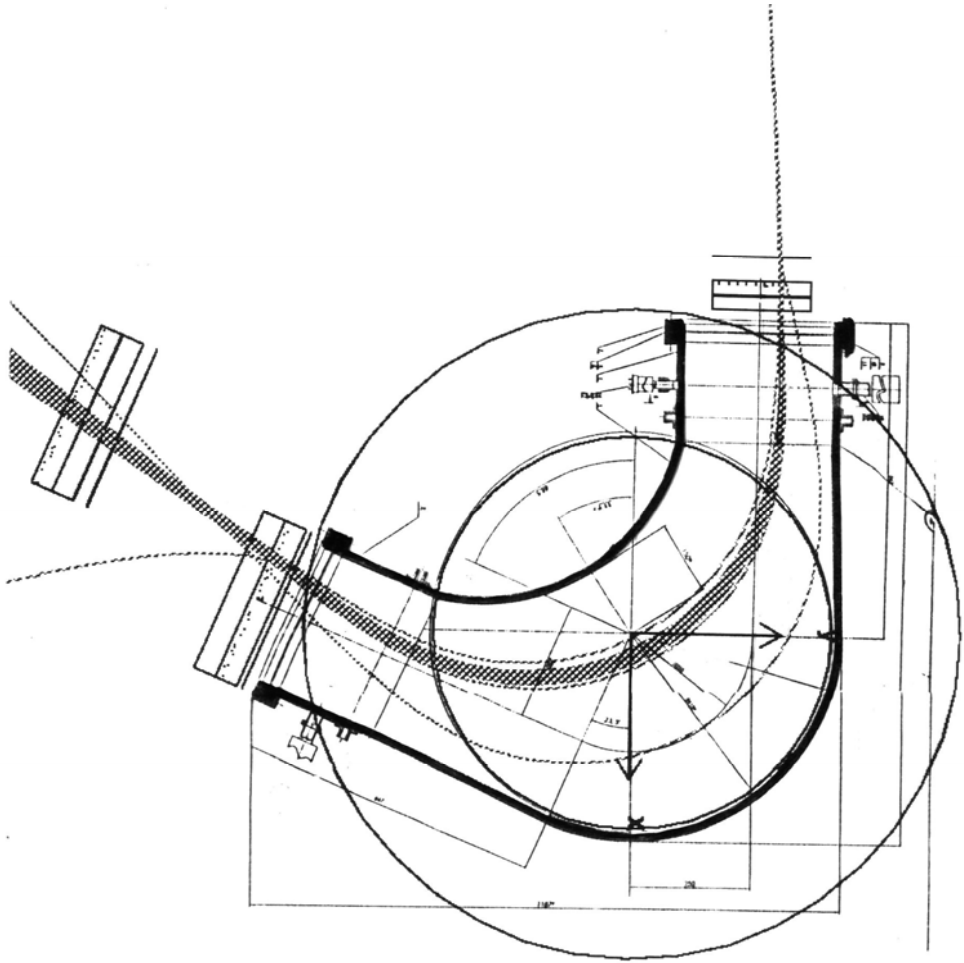

Fig.5. The simulation results for 52.8 MeV positron trajectories for FAMILON setup

Note, that the real momentum precision will be slightly higher because only one chamber coordinates have been used in the above evaluations, but for the momentum reconstruction at least two chambers are used.

The further improvement of the precision in the momentum measurement may be obtained only if the constructive improvement of the coordinate detectors will decrease thequantity of substance on the positron path.

The important parameter of the experiment is the efficiency of the positron registration as a function of positron energy and the exit angle. This efficiency also was calculated by means of Monte Carlo-method for the optimal setup configuration. It was assumed that positrons go isotropically out of the target in the angle up to 8° and the positron spectrum is described by Mishel function. The spectrum in the interval 35 – 52.8 Mev was considered. The efficiency as a function of two parameters – the positron momentum and the angle out of the target – was determined as the ratio of two-dimensional histograms: one for registered positrons in chambers PC1 – PC4, another – for positrons going out of the target. Fig.6 demonstrates the efficiency dependence on the positron momentum for various exit angles (0° – solid line, 2° – dashed line, 4° – dotted line, 6° – dashed-dotted line, 8° – solid line). It can be seen from fig.6 that the registration efficiency for positrons in the energy interval from 45 MeV up to maximal value does not fall less than 0.5 up to 6°.

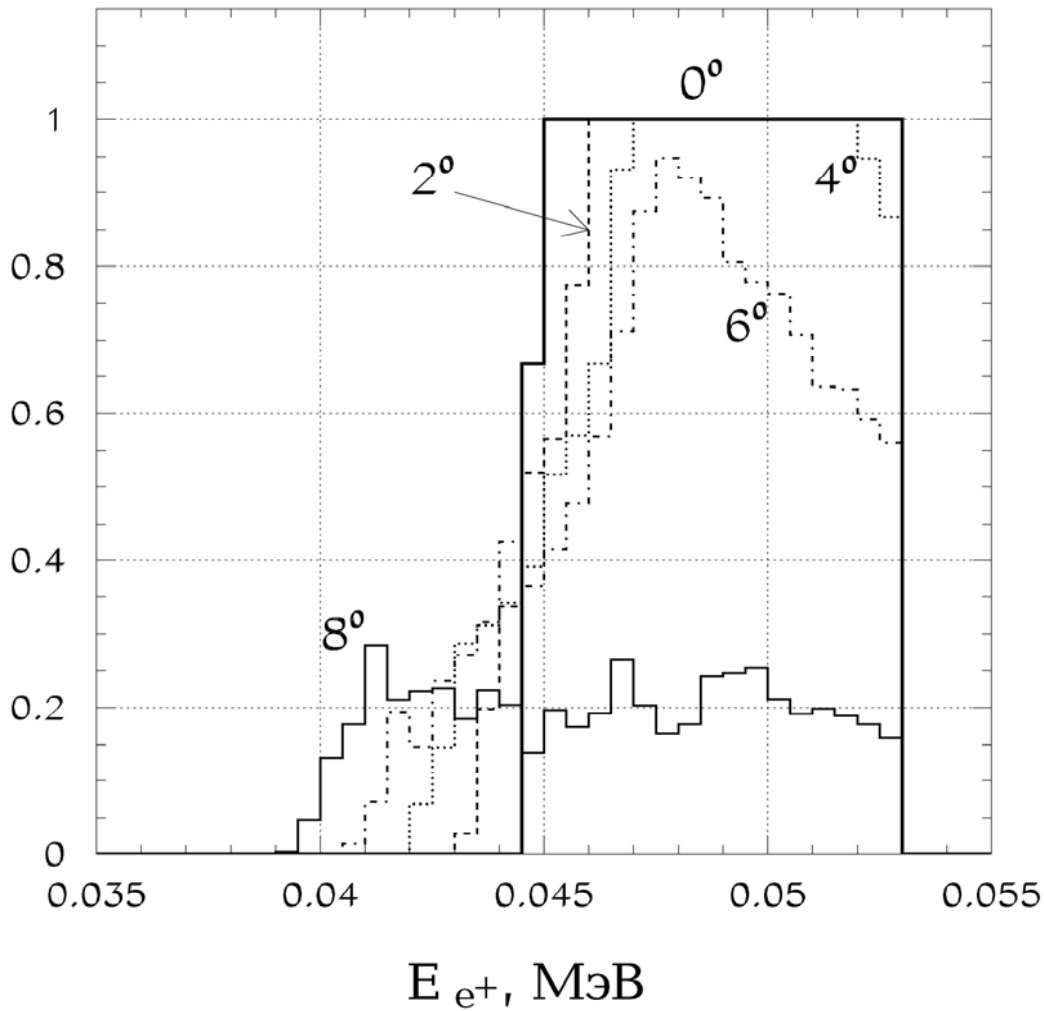

Fig.6. The dependence of the positron registration efficiency on its energy for five exit angles (0° – solid line, 2° – dashed line, 4° – dotted line, 6° – dashed-dotted line, 8° – solid line).

# 6. Conclusions

Lepton family violation processes belong to a few experimentally accessible signatures of new physics predicted by particle theory. The mode of decay with light boson is studied at much lower level of experimental sensitivity, than the other types of lepton family violation transitions, such as µ→e+ light boson. On the other hand, the theoretical models of family symmetry breaking favour free familon emission modes as much more sensitive probe for such models, than the second order FCNC processes (like µ→3e or µ→eγ). So the progress in the experimental situation leads to the substantial progress in studying new physics. The cosmological and astrophysical arguments make it possible to adjust the parameters of familons models and to specify the expected rate for familon decays. The search for such decays turns to be an *experimentum crucis* for dark matter cosmological scenarios and familon models, underlying them. Thus the progress in the search for µ →eα decay is important for both particle physics and cosmology.

The experiment "FAMILON" is carrying out in the frameworks of special scietific-tecnical Russia Program "Research and elaboration in priority directions of scientific and technical development for civil purposes", subprogram "Fundamental nuclear physics" project "Physics of rare processes" (State contract № 40.052.1.1.1111 on 31.01.2003); also in the frameworks of RAS Presidium Program "Neutrino physics" (State contract № 10002-251/П-06/048-058/100603-602 on 01.04.2003), RPPS Project (project 99-02-17943-a) and Program of support the leading science schools in Russia (project SS-1867.2003,2).